\documentclass[runningheads]{llncs}
\usepackage[T1]{fontenc}
\usepackage{graphicx}
\usepackage{wrapfig}
\usepackage{caption}
\usepackage{cite}
\usepackage{amsmath,amssymb,amsfonts}
\usepackage{algorithmic}
\usepackage{graphicx}
\usepackage{textcomp}
\usepackage{xcolor}
\usepackage{booktabs}
\usepackage{url}
\usepackage{multirow}
\usepackage{enumitem}
\usepackage{dsfont}
\usepackage[caption=false,font=footnotesize]{subfig} 
\usepackage[most]{tcolorbox}
\usepackage{tabularx}
\usepackage{ragged2e}
\usepackage{ulem}
\tcbset{
   colback=white, colframe=black!15, coltitle=black,
   boxrule=0.4pt, arc=1mm, left=1.2ex, right=1.2ex, top=0.8ex, bottom=0.8ex,
   enhanced, sharp corners=all
}
\usepackage{marvosym}
\usepackage[table]{xcolor}
\newcommand{\best}[1]{\textbf{#1}}
\newcommand{\secondbest}[1]{\underline{#1}}
\newcommand{\worst}[1]{\dashuline{#1}}
\newcommand{\secondworst}[1]{#1}
\newcommand{\sd}[1]{\scalebox{0.8}{$\pm$#1}}

\newcommand\bk[1]{\textcolor{black}{#1}}

\begin{document}
\title{On the Factual Consistency of Text-based Explainable Recommendation Models}

\titlerunning{On the Factual Consistency of Text-based Explainable Recommendation}

\author{Ben Kabongo\inst{1}$^\text{\Letter}$
\and
Vincent Guigue\inst{2}}
\authorrunning{B. Kabongo and V. Guigue}
\institute{Sorbonne University, CNRS, ISIR, Paris, France \and
AgroParisTech, UMR MIA Paris-Saclay, Palaiseau, France \\
\email{ben.kabongo@sorbonne-universite.fr}, 
\email{vincent.guigue@agroparistech.fr}}

\maketitle      

\begin{abstract}
Text-based explainable recommendation aims to generate natural-language explanations that justify item recommendations, to improve user trust and system transparency.
Although recent advances leverage LLMs to produce fluent outputs, a critical question remains underexplored: are these explanations factually consistent with the available evidence?
We introduce a comprehensive framework for evaluating the factual consistency of text-based explainable recommenders.
We design a prompting-based pipeline that uses LLMs to extract atomic explanatory statements from reviews, thereby constructing a ground truth that isolates and focuses on their factual content.
Applying this pipeline to five categories from the Amazon Reviews dataset, we create augmented benchmarks for fine-grained evaluation of explanation quality.
We further propose statement-level alignment metrics that combine LLM- and NLI-based approaches to assess both factual consistency and relevance of generated explanations.
Across extensive experiments on six state-of-the-art explainable recommendation models, we uncover a critical gap: while models achieve high semantic similarity scores (BERTScore F1: 0.81-0.90), all our factuality metrics reveal alarmingly low performance (LLM-based statement-level precision: 4.38\%-32.88\%).
These findings underscore the need for factuality-aware evaluation in explainable recommendation and provide a foundation for developing more trustworthy explanation systems.~\footnote{\bk{Code, datasets and illustrations are available at \url{https://github.com/BenKabongo25/factual_explainable_recommendation}.
}}

\end{abstract}

\section{Introduction}
Recommender systems have become integral to modern digital platforms, guiding users through vast catalogs of products, content, and services. However, traditional recommendation approaches~\cite{he2020lightgcn,koren2009matrix} often operate as black boxes, providing little insight into \textit{why} a particular item is suggested.
Explainable recommendation addresses this limitation by generating human-interpretable justifications for recommendations, thereby enhancing transparency, user satisfaction, and trust~\cite{zhang2020explainable}.

\begin{wrapfigure}{r}{0.45\linewidth}
\vspace{-0.5\baselineskip} 
\resizebox{1\linewidth}{!}{
\includegraphics[width=1\linewidth]{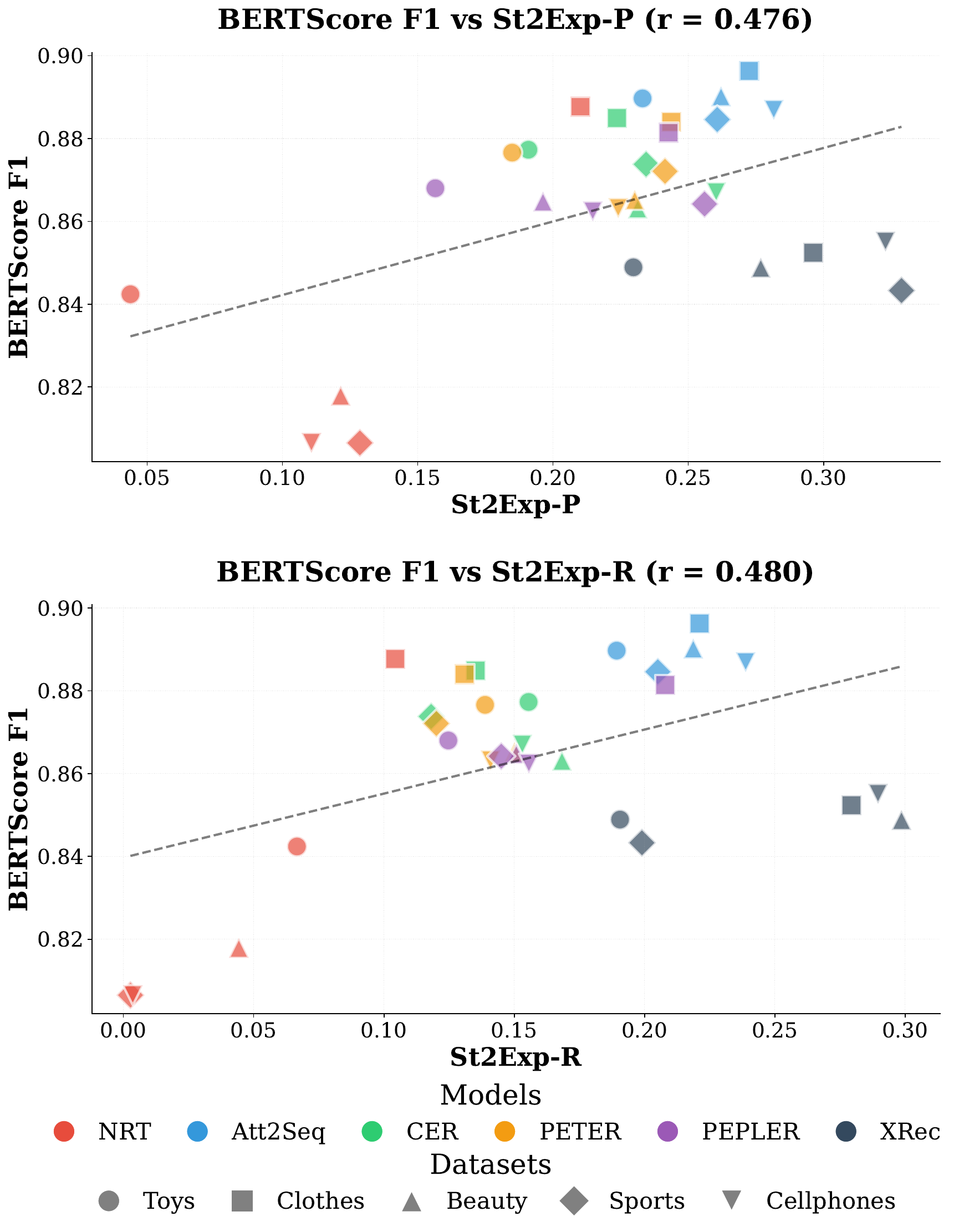}
}
\captionof{figure}{BERTScore F1 vs. LLM-based statement-level metrics Precision (top) and Recall (bottom). Pearson correlation (r) reported in brackets.}
\label{fig:bertscore_vs_st2exp_comparison}
\vspace{-0.5\baselineskip}
\end{wrapfigure}

\noindent
Among various explanation paradigms, text-based explainable recommendation has emerged as a particularly promising approach, leveraging the flexibility and expressiveness of natural language to convey personalized rationales~\cite{dong2017learning,li2017neural,li2021personalized,li2023personalized,li2025g,ma2024xrec}. Recent advances have increasingly turned to LLMs~\cite{achiam2023gpt,dubey2024llama}, which can generate fluent and contextually appropriate explanations. 
These models have demonstrated impressive performance on standard text generation metrics~\cite{fu2023gptscore,sellam2020bleurt,yuan2021bartscore,zhang2019bertscore}, producing explanations that appear coherent and plausible at first glance.

Yet a critical question remains largely unexplored: \textit{Are the explanations generated by state-of-the-art models factually consistent with the available evidence?} 
While surface-level fluency and lexical matching are important, the true value of an explanation lies in its \textit{factuality}: whether the explanatory content accurately reflects the user's actual preferences as expressed in their reviews. 
We introduce a comprehensive framework for evaluating the factual consistency of text-based explainable recommenders, comprising four key contributions:

\noindent
\textbf{(1) Statement-Level Ground-truth.} We design a prompting-based pipeline that uses LLMs to extract atomic explanatory statements from user reviews with their associated domain-specific topics and sentiment labels.
A rule-based aggregation procedure then constructs factual ground-truth explanations. 

\noindent
\textbf{(2) Augmented Benchmark Datasets.} We apply our pipeline to five Amazon Reviews categories~\cite{ni2019justifying} (\textit{Toys and Games}, \textit{Clothing}, \textit{Beauty}, \textit{Sports}, and \textit{Cellphones}) 
creating augmented benchmarks pairing each user–item interaction with extracted statements and derived ground-truth explanations.

\noindent
\textbf{(3) Statement-Level Factuality Metrics.} Building on recent advances~\cite{herserant2025seval,laban2022summac,zha2023alignscore}, we propose metrics tailored to explainable recommendation. Combining LLM-based and NLI approaches, our metrics assess factual consistency (precision and recall) at the statement level.

\noindent
\textbf{(4) Experiments.} We conduct extensive experiments on six state-of-the-art explainable recommendation models spanning three architectural families: recurrent models (NRT~\cite{li2017neural}, Att2Seq~\cite{dong2017learning}), transformer-based models (PETER~\cite{li2021personalized}, CER~\cite{raczynski2023problem}, PEPLER~\cite{li2023personalized}), and LLM-enhanced models (XRec~\cite{ma2024xrec}). 

Our evaluation reveals substantial gaps between surface-level text quality and factual accuracy. While models achieve high scores on standard similarity metrics (BERTScore F1 from 0.81 to 0.90), our statement-level metrics tell a different story (Figure~\ref{fig:bertscore_vs_st2exp_comparison}). In particular, LLM-based precision (\texttt{St2Exp-P} from 4.38\% to 32.88\%) and recall (\texttt{St2Exp-R} from 0.27\% to 29.86\%) are very low. This discrepancy highlights the limitations of existing evaluation practices and underscores the prevalence of hallucinated explanatory content in current systems.
Our findings have important implications for the design and evaluation of explainable recommenders. 
Addressing this challenge will require not only better evaluation methodologies, such as the statement-level metrics we propose, but also fundamental innovations in model architectures and training objectives that explicitly prioritize factual grounding.

\section{Related Work}

\noindent
\textbf{Text-based Explainable Recommendation.}
Text-based explainable recommendation~\cite{dong2017learning,kabongo2025elixir,li2017neural,li2021personalized,li2023personalized,li2025g,ma2024xrec,xie2023factual} has gained momentum, especially in combination with language models, producing tips~\cite{li2017neural}, reviews~\cite{dong2017learning,li2021personalized,li2023personalized,xie2023factual}, or an explanatory paragraph derived from the review~\cite{li2025g,ma2024xrec}.
\textit{(1) RNN-based approaches} include Att2Seq~\cite{dong2017learning} and  NRT~\cite{li2017neural}.
\textit{(2) Transformer-based approaches} include PETER~\cite{li2021personalized} with an untrained Transformer~\cite{vaswani2017attention}, extended by CER~\cite{raczynski2023problem} to align explanations with ratings. PEPLER~\cite{li2023personalized} fine-tunes GPT-2~\cite{radford2019language} for explanation generation.
\textit{(3) LLM-enhanced approaches} incorporate collaborative-filtering signals via GNNs~\cite{he2020lightgcn}. XRec~\cite{ma2024xrec} injects latent representations into LLMs through a collaborative adapter, while G-Refer~\cite{li2025g} uses a hybrid graph-retrieval mechanism for explanation generation.
\\

\noindent
\textbf{Evaluation of Text-based Explainable Recommenders.}
Early work largely relied on n-gram overlap metrics such as BLEU~\cite{papineni2002bleu} and ROUGE~\cite{lin2004rouge}. 
However, these metrics depend on exact word matching and n-gram overlap, which makes them struggle with synonymy and paraphrasing.
More recent studies therefore incorporate semantics-aware metrics, including GPTScore~\cite{fu2023gptscore}, BERTScore~\cite{zhang2019bertscore}, BARTScore~\cite{yuan2021bartscore}, and BLEURT~\cite{sellam2020bleurt}. 
While useful for comparing systems on surface-level generation quality, these metrics do not adequately capture the factuality of generated explanations~\cite{honovich2022true,laban2022summac,zha2023alignscore}. 
\\

\noindent
\textbf{Factual Consistency Evaluation.}
Factual consistency is critical in summarization~\cite{zha2023alignscore,honovich2022true} and LLM evaluation~\cite{huang2024trustllm,min2023factscore}. Early n-gram metrics~\cite{banerjee2005meteor,lin2004rouge,papineni2002bleu} showed weak correlation with human judgments. Later approaches used NLI~\cite{laban2022summac,zha2023alignscore}, LLMs~\cite{herserant2025seval}, or QA~\cite{scialom2021questeval}, but decomposed texts into coarse chunks limiting fine-grained evaluation of recommendation explanations.
Metrics like FMR, FCR, and DIV~\cite{li2020generate} assess item feature consistency via exact matching, ignoring sentiment. \cite{xie2023factual} employs entailment but omits ground-truth comparison. We propose statement-level metrics combining NLI and LLM scoring for fine-grained factuality assessment in explainable recommendation.

\section{Framework for Factual Explainable Recommendation}
User reviews typically combine explanatory content that justifies the rating with noise, including irrelevant, non-explanatory text.
Depending on the domain (e.g., clothing), explanatory content can be grouped into high-level topics (e.g., \textit{fit, material}).
The goal of text-based explainable recommendation is to jointly predict the user’s rating for the item and generate a textual rationale that explains the underlying user–item interaction.
In this section, we describe our dataset construction pipeline and the metrics used to evaluate the factual consistency of text-based explainable recommenders.

\begin{wrapfigure}{r}{0.45\linewidth}
  \vspace{-3mm} 
  \captionof{table}{Ground-truth construction illustration (\textit{explanatory statements} are shown in \textit{italic}).}
  \label{tab:gt-const-illustration}
  \vspace{-3mm} 
  \centering
\resizebox{1\linewidth}{!}{
\begin{tcolorbox}[width=1.2\linewidth]
\footnotesize
\begin{tabularx}{\linewidth}{@{}>{\RaggedRight\arraybackslash}X@{}}
\textbf{Review}: Got this sweater as a gift for my sister. \textcolor{red}{\textit{The material feels cheap}}. I've been shopping with this brand for years now. \textcolor{orange}{\textit{It runs true to size}}. \textcolor{blue}{\textit{The design is really cute}} though! Would recommend checking their other products.\par\medskip
\textbf{Ground-truth Explanation}: The user would appreciate this product because \textcolor{blue}{\textit{it has a really cute design}}. However, they may dislike that \textcolor{red}{\textit{the material feels cheap}}. They seem indifferent to \textcolor{orange}{\textit{it runs true to size}}.\par\medskip
\end{tabularx}
\end{tcolorbox}
}
\label{tab:ground-truth-explanation}
\vspace{-5mm} 
\end{wrapfigure}
\subsection{Statement Extraction and Ground-truth Construction}
\label{sec:sts-extraction}
An \textit{atomic explanatory statement} is a polarized fact expressing the user’s opinion about a single attribute or topic of the item.
Given the review $\mathbf{t}_{ui}$ written by user $u$ for item $i$, our first goal is to extract the atomic explanatory statements together with their corresponding topics and polarities.
Topics enable pre-filtering of non-explanatory noise in reviews.
To preserve the full explanatory content extracted from each review, we construct an explanatory paragraph from the triplets using a rule-based procedure, as shown in Table~\ref{tab:gt-const-illustration}.
\\

\noindent
\textbf{Statement Extraction.}
For each domain, we specify a set of topics of interest~$\mathcal{T}$. 
In our experiments, we first design a prompt to elicit a shortlist of domain-specific topics.
Given the topic set $\mathcal{T}$ and the sentiment label set $\mathcal{P} = \{\text{\textsc{pos}}, \text{\textsc{neg}}, \text{\textsc{neu}} \}$, we design a domain-specific prompt to extract from $\mathbf{t}_{ui}$ the set of atomic statements and assign to each statement its topic and polarity, as follows:
\begin{align}
    \label{eq:statement-triplets}
    S_{ui} 
    = \mathrm{LLM}(\mathbf{t}_{ui} \mid \cdot, \mathcal{T}, \mathcal{P})
    = \{(\mathbf{s}_1, t_1, p_1), \cdots, (\mathbf{s}_{n_{ui}}, t_{n_{ui}}, p_{n_{ui}})\},
\end{align}
where $\mathbf{s}_k$ is the $k$-th atomic explanatory statement in $\mathbf{t}_{ui}$, $t_k \in \mathcal{T}$ is its topic, and $p_k \in \mathcal{P}$ is its sentiment label.
\\

\noindent
\textbf{Ground-truth Explanation Construction.}
After obtaining the triplet set $S_{ui}$, we apply a rule-based procedure to compose a single explanatory paragraph that combines all statements.
We first group statements by polarity; for each polarity present, we form a sentence that aggregates its associated statements.
%
%
%
We then concatenate the resulting sentences with simple logical connectors to form a well-structured paragraph.
This approach avoids the cost of using an LLM for this step and ensures that all extracted statements are preserved in the constructed explanation.
Table~\ref{tab:gt-const-illustration} provides an illustrative example of ground-truth construction.

\subsection{Datasets}
\label{sec:datasets}
\begin{wrapfigure}{r}{0.52\linewidth}
  \vspace{-2mm} 
  \captionof{table}{Dataset statistics.}
  \label{tab:datasets}
  \vspace{-0.5\baselineskip} 
  \centering
\resizebox{1\linewidth}{!}{
\setlength{\tabcolsep}{4pt}
\begin{tabular}{@{}lrrrrr@{}}
\toprule
 & 
\textbf{Toys} & 
\textbf{Clothes} & 
\textbf{Beauty} &
\textbf{Sports} & 
\textbf{Cell}\\
\midrule
\textbf{Users} & 19 398 & 39 385 & 22 362 & 35 596 & 27 873\\
\textbf{Items} & 11 924 & 23 033 & 12 101 & 18 357 & 10 429\\
\textbf{Interactions} & 163 711 & 274 774 & 197 621 & 293 244 & 190 194\\
Train & 121 751 & 203 574 & 149 569 & 219 913 & 139 889\\
Validation & 14 805 & 24 396 & 18 506 & 27 394 & 16 099\\
Test & 22 441 & 41 995 & 27 862 & 42 675 & 28 901\\
\midrule 
\textbf{Statements} \\
Avg/interaction & 5.03 & 4.42 & 5.45 & 4.93 & 4.54\\
Avg/user & 41.76 & 30.12 & 46.99 & 40.24 & 30.65\\
Avg/item & 67.49 & 50.70 & 84.79 & 76.90 & 81.42\\
Unique & 587 114 & 619 917 & 622 276 & 1 055 145 & 662 466\\
Total  & 823 932 & 1 215 270 & 1 076 769 & 1 447 240 & 863 036\\
\bottomrule
\end{tabular}
}
\label{tab:datasets}
\end{wrapfigure}
We apply our pipeline to five categories from the Amazon Reviews 2014 dataset\footnote{\url{https://jmcauley.ucsd.edu/data/amazon/links.html}}~\cite{ni2019justifying}.
The categories are \textbf{Toys and Games} (\textit{Toys}), \textbf{Clothing, Shoes and Jewelry} (\textit{Clothes}), \textbf{Beauty} (\textit{Beauty}), \textbf{Sports and Outdoors} (\textit{Sports}), and \textbf{Cell Phones and Accessories} (\textit{Cellphones}).
In our experiments, we used \texttt{Llama-3-8B-Instruct}\footnote{\url{https://huggingface.co/meta-llama/Meta-Llama-3-8B-Instruct}}~\cite{dubey2024llama}.
For each dataset, we define a list of ten topics of interest and design a dataset-specific prompt with a few illustrative examples to extract all explanatory triplets from every interaction. From each review and its extracted statements, we then construct the corresponding ground-truth explanation.
Dataset statistics are reported in Table~\ref{tab:datasets}.

\subsection{Evaluation Metrics}
\label{sec:evaluation-metrics}

A good explainable recommender should generate explanations whose passages are all supported by the reference (precision) while also covering as many of the reference’s explanatory passages as possible (recall).
Accordingly, we derive a set of metrics to measure the factual consistency (precision and recall) of a generated explanation at the statement level.
Given $m$ statements $\{\mathbf{s}_1,\ldots,\mathbf{s}_m\}$ extracted from the ground-truth explanation $\mathbf{e}$ and $n$ statements $\{\mathbf{s}'_1,\ldots,\mathbf{s}'_n\}$ extracted from the generated explanation $\mathbf{e}'$, we introduce two families of statement-level metrics: LLM-based and NLI-based metrics.

\vspace{-5mm}
\subsubsection{LLM-based Metrics.}
These metrics use an LLM-based scoring function $f_{\mathrm{LLM}}$ to assess the factual consistency of a statement given a target text unit. 
We define: \textit{Statement-to-Explanation Precision} (\texttt{St2Exp-P}), \textit{Statement-to-Explanation Recall} (\texttt{St2Exp-R}), and \textit{Statement-to-Explanation F1} (\texttt{St2Exp-F1}).
All follow the SEval~\cite{herserant2025seval} framework, and are given by:
\begin{align}
    \texttt{St2Exp-P} \;=\; \frac{1}{n} &\sum_{k=1}^{n} f_{\mathrm{LLM}}(\mathbf{s}'_k, \mathbf{e}), \quad
    \texttt{St2Exp-R} \;=\; \frac{1}{m} \sum_{l=1}^{m} f_{\mathrm{LLM}}(\mathbf{s}_l, \mathbf{e}'),  \notag \\ 
    &\texttt{St2Exp-F1} \;=\;2\; \frac{\,\texttt{St2Exp-P}\cdot \texttt{St2Exp-R}}{\texttt{St2Exp-P} + \texttt{St2Exp-R}}.
\end{align}

\subsubsection{NLI-based Metrics.}
Our NLI-based metrics use an entailment-based scoring function $f_{\mathrm{NLI}}$ to evaluate factual consistency between pairs of statements.
We define two subgroups of metrics, differing in the choice of $f_{\mathrm{NLI}}(\mathbf{s}_k, \mathbf{s}_l)$ for statements $\mathbf{s}_k$ and $\mathbf{s}_l$.
In the first subgroup, the score is the entailment probability $E_{kl}$; we denote this by $f_{\mathrm{NLI\text{-}ent}}$.
In the second subgroup, the score is the difference between entailment and contradiction, $E_{kl} - C_{kl}$; we refer to this as the \textit{coherence score} and denote it by $f_{\mathrm{NLI\text{-}coh}}$.
Note that these measures are not symmetrical.
We define precision-oriented metrics (\texttt{StEnt-P} and \texttt{StCoh-P}), recall-oriented metrics (\texttt{StEnt-R} and \texttt{StCoh-R}) and a harmonic mean metric (\texttt{StEnt-F1}).
These metrics are defined as follows:
\begin{align}
    \texttt{St*-P} \;=\; \frac{1}{n} \sum_{k=1}^{n} \max_{l}\,& f_{\mathrm{NLI\text{-}*}}(\mathbf{s}'_k, \mathbf{s}_l), \quad
    \texttt{St*-R} \;=\; \frac{1}{m} \sum_{l=1}^{m} \max_{k}\, f_{\mathrm{NLI\text{-}*}}(\mathbf{s}_l, \mathbf{s}'_k), \notag \\
    &\texttt{St*-F1} \;=\; 2\,\frac{\texttt{St*-P}\cdot\texttt{St*-R}}{\texttt{St*-P} + \texttt{St*-R}},
\end{align}
where $*\in\{\text{ent},\text{coh}\}$ indexes the chosen scoring function.

\section{Experiments}
\subsection{Experimental Protocol}

\noindent
\textbf{Baselines.}
We consider three families of state-of-the-art models in our evaluation: RNN–based models (Att2Seq~\cite{dong2017learning} and NRT~\cite{li2017neural}), Transformer-based models (PETER~\cite{li2021personalized}, CER~\cite{raczynski2023problem} and PEPLER~\cite{li2023personalized}), and LLM-based models (XRec~\cite{ma2024xrec}).
For XRec~\cite{ma2024xrec} we use \texttt{Llama-2-7b}\footnote{\url{https://huggingface.co/meta-llama/Llama-2-7b-chat-hf}}~\cite{touvron2023llama}, following the original paper.
For each model, we adopt the best hyperparameters reported in the corresponding paper.
\\

\noindent
\textbf{Datasets.}
We evaluate on five categories from the Amazon Reviews dataset~\cite{ni2019justifying}: \textit{Toys}, \textit{Clothes}, \textit{Beauty}, \textit{Sports}, and \textit{Cellphones}.
Dataset construction details and statistics are provided in Section~\ref{sec:datasets}.
\\

\noindent
\textbf{Evaluation Metrics.}
We evaluate all models with the suite introduced in Section~\ref{sec:evaluation-metrics}. 
We employ \texttt{Llama-3.1-8B-Instruct}\footnote{\url{https://huggingface.co/meta-llama/Llama-3.1-8B-Instruct}}~\cite{dubey2024llama} for our LLM-based metrics and \texttt{DeBERTa-large-mnli}\footnote{\url{https://huggingface.co/microsoft/deberta-large-mnli}}~\cite{he2021deberta} for our NLI-based metrics.
For all models, we train on the training split, use the validation split for model selection, and report, for each metric, the mean and standard deviation computed over all test examples.

\begin{figure*}[htbp]
\centerline{\includegraphics[width=\linewidth]{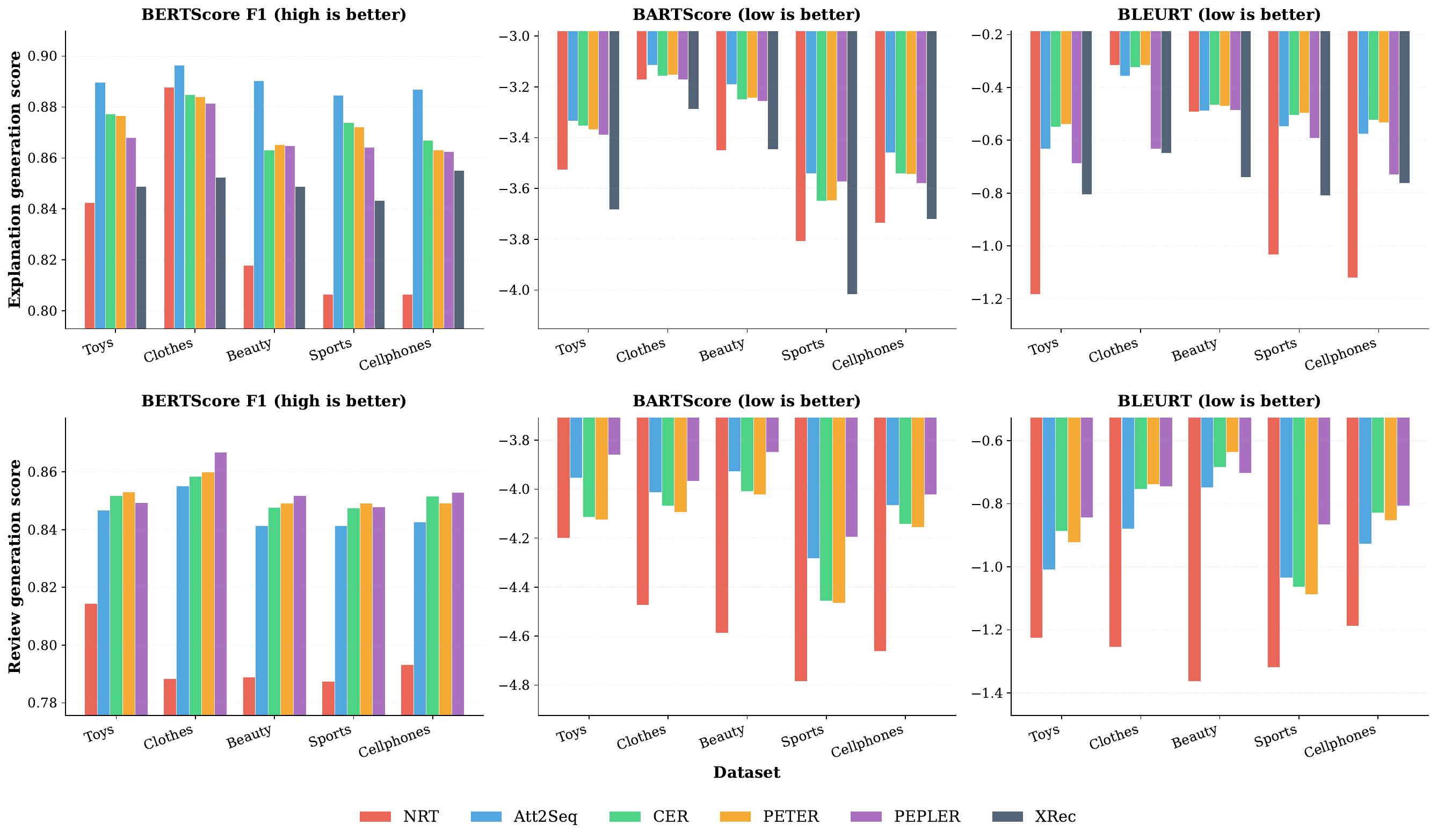}}
\caption{Text Similarity Results on explanation generation (top) and review generation (bottom).}
\label{fig:text-similarity-results}
\vspace{-5mm}
\end{figure*}
\subsection{Text Similarity Results}

\noindent
\textbf{Explanation Generation.}
Figure~\ref{fig:text-similarity-results} (top) presents model evaluation using similarity metrics common in recent explainable recommendation work, with models evaluated against statement-derived ground-truth explanations.
Results show trends contrary to prior work evaluating on shorter explanations. XRec ranks among the lowest on similarity metrics, outperformed even by NRT on \textit{Clothes}. Att2Seq performs best across datasets on most metrics, followed closely by CER, while PEPLER consistently achieves the highest BLEURT scores.
\\

\noindent
\textbf{Review Generation.}
Figure~\ref{fig:text-similarity-results} (bottom) reports model performance (excluding XRec) on review generation using text similarity metrics. Results align with those reported in PEPLER and PETER (evaluated on review generation with n-gram metrics), confirming that Transformer-based models generate reviews more similar to actual reviews than RNN-based predecessors, though this does not guarantee factuality.
The disparity between explanation and review results suggests models excel at either explanation or review generation, but not necessarily both. However, since reviews contain noise alongside explanatory content, neither high review nor explanation similarity guarantees factual generation.

\begin{table*}[htbp]
\vspace{-6mm}
\caption{LLM-based Statement-Level Evaluation Results (\best{best}, \secondbest{second}, \worst{worst}).}
\setlength{\tabcolsep}{3pt}
\begin{center}
\vspace{-2mm}
\resizebox{0.8\linewidth}{!}{
\begin{tabular}{lcccccc}
\toprule
& 
\textbf{NRT} & \textbf{Att2Seq} & \textbf{CER} & \textbf{PETER} & \textbf{PEPLER} & \textbf{XRec} \\
\midrule
\multicolumn{7}{c}{\textbf{Toys}} \\
\midrule
  \text{St2Exp-P} & \worst{0.0438\sd{0.1772}} & \best{0.2331\sd{0.2844}} & 0.1909\sd{0.2793} & 0.1849\sd{0.2827} & \secondworst{0.1565\sd{0.2731}} & \secondbest{0.2297\sd{0.3039}} \\
  \text{St2Exp-R} & \worst{0.0666\sd{0.1441}} & \secondbest{0.1893\sd{0.2498}} & 0.1555\sd{0.2371} & 0.1388\sd{0.2245} & \secondworst{0.1247\sd{0.2053}}  & \best{0.1906\sd{0.2617}} \\
  \text{St2Exp-F1} & \worst{0.0091\sd{0.0597}} & \best{0.1317\sd{0.1954}} & 0.0988\sd{0.1772} &  0.0917\sd{0.1736} & \secondworst{0.0728\sd{0.1550}} & \secondbest{0.1124\sd{0.1887}} \\
\midrule
\multicolumn{7}{c}{\textbf{Clothes}} \\
\midrule
  \text{St2Exp-P} & \worst{0.2102\sd{0.2482}} & \secondbest{0.2725\sd{0.2998}} & \secondworst{0.2236\sd{0.2935}} & 0.2438\sd{0.3096} & 0.2427\sd{0.2639} & \best{0.2962\sd{0.2897}} \\
  \text{St2Exp-R} & \worst{0.1044\sd{0.1706}} & \secondbest{0.2211\sd{0.2646}} & 0.1351\sd{0.2128} & \secondworst{0.1309\sd{0.2117}} & 0.2079\sd{0.2466} & \best{0.2794\sd{0.3003}} \\
  \text{St2Exp-F1} & \worst{0.0830\sd{0.1465}} & \secondbest{0.1613\sd{0.2149}} & \secondworst{0.0982\sd{0.1746}} & 0.0987\sd{0.1775} & 0.1521\sd{0.2016} & \best{0.1913\sd{0.2257}} \\
\midrule
\multicolumn{7}{c}{\textbf{Beauty}} \\
\midrule
  \text{St2Exp-P} & \worst{0.1215\sd{0.3267}} & \secondbest{0.2621\sd{0.3020}} & 0.2313\sd{0.3281} & 0.2302\sd{0.3261} & \secondworst{0.1963\sd{0.2860}} & \best{0.2768\sd{0.3068}} \\
  \text{St2Exp-R} & \worst{0.0443\sd{0.1135}} & \secondbest{0.2187\sd{0.2557}} & 0.1683\sd{0.2291} & \secondworst{0.1501\sd{0.2207}} & 0.1508\sd{0.2077} & \best{0.2986\sd{0.3188}} \\
  \text{St2Exp-F1} & \worst{0.0391\sd{0.1277}} & \secondbest{0.1552\sd{0.2128}} & 0.1203\sd{0.1978} & 0.1102\sd{0.1960} & \secondworst{0.0999\sd{0.1751}} & \best{0.1735\sd{0.2266}} \\
\midrule
\multicolumn{7}{c}{\textbf{Sports}} \\
\midrule
  \text{St2Exp-P} & \worst{0.1286\sd{0.3347}} & \secondbest{0.2607\sd{0.2887}} & \secondworst{0.2344\sd{0.3423}} & 0.2414\sd{0.3491} & 0.2560\sd{0.3156} & \best{0.3288\sd{0.3663}} \\
  \text{St2Exp-R} & \worst{0.0027\sd{0.0269}} & \best{0.2051\sd{0.2512}} & \secondworst{0.1181\sd{0.1979}} & 0.1201\sd{0.1997} & 0.1450\sd{0.2145} & \secondbest{0.1990\sd{0.2626}} \\
  \text{St2Exp-F1} & \worst{0.0037\sd{0.0377}} & \best{0.1511\sd{0.2053}} & \secondworst{0.0929\sd{0.1810}} & 0.0958\sd{0.1848} & 0.1123\sd{0.1853} & \secondbest{0.1473\sd{0.2216}} \\
\midrule
\multicolumn{7}{c}{\textbf{Cellphones}} \\
\midrule
  \text{St2Exp-P} & \worst{0.1107\sd{0.2981}} & \secondbest{0.2816\sd{0.3033}} & 0.2603\sd{0.3435} & 0.2241\sd{0.3297} & \secondworst{0.2147\sd{0.2993}} & \best{0.3229\sd{0.3369}} \\
  \text{St2Exp-R} & \worst{0.0036\sd{0.0351}} & \secondbest{0.2388\sd{0.2777}} & 0.1531\sd{0.2327} & \secondworst{0.1409\sd{0.2262}} & 0.1556\sd{0.2246} & \best{0.2896\sd{0.3301}} \\
  \text{St2Exp-F1} & \worst{0.0005\sd{0.0129}} & \secondbest{0.1679\sd{0.2206}} & 0.1169\sd{0.1990} & \secondworst{0.1027\sd{0.1891}} & 0.1071\sd{0.1849} & \best{0.1825\sd{0.2430}} \\
\bottomrule
\end{tabular}
}
\label{tab:llm-statement-level-results}
\end{center}
\vspace{-6mm}
\end{table*}
\subsection{LLM-based Statement Evaluation (St2Exp)}
Table~\ref{tab:llm-statement-level-results} reports results on our LLM-based factual-consistency metrics.
Across models and datasets, scores are uniformly low for both precision and recall.
\\

\noindent
\textbf{Precision.}
The highest average precision is achieved by XRec on \textit{Sports} with $32.88\%$ (std.\ $33.89\%$); despite being the best, this remains modest and can be problematic in settings where precision is critical. Overall, \texttt{St2Exp-P} indicates that state-of-the-art systems exhibit low precision in explanation generation. NRT yields the lowest precision, e.g., $4.38\%$ on \textit{Toys}, and performs poorly across datasets and metrics.
\\

\noindent
\textbf{Recall.} 
\texttt{St2Exp-R} is generally even lower than precision, confirming that current models fail to recover most of the ground-truth explanatory passages. This shortfall is especially concerning in scenarios that require comprehensive coverage or when certain passages are particularly salient to users. XRec attains the best recall at $29.86\%$ on \textit{Beauty}. 
The lowest recall is observed for NRT with $0.27\%$ on \textit{Sports}.
Finally, \texttt{St2Exp-F1} corroborates these findings, with values ranging from $0.05\%$ (NRT on \textit{Cellphones}) to $19.13\%$ (XRec on \textit{Clothes})
\\

\noindent
\textbf{Comparison with text similarity metrics.}
Critically, comparing text similarity metrics with our LLM-based factuality metrics reveals a striking disconnect: despite low factual precision and recall scores, models achieve very high similarity scores, with BERTScore F1 ranges from 0.81 to 0.90 (Figure~\ref{fig:bertscore_vs_st2exp_comparison}).
This dramatic gap demonstrates that models can be semantically similar while being factually inconsistent: they use similar vocabulary and phrasing while making unsupported or contradictory claims.
This underscores the need to revise experimental protocols for factuality-oriented models and develop robust, adapted metrics that properly evaluate factual consistency.

\begin{table*}[htbp]
\vspace{-6mm}
\caption{NLI-based Statement-Level Evaluation Results  (\best{best}, \secondbest{second}, \worst{worst}).}
\setlength{\tabcolsep}{3pt}
\vspace{-2mm}
\begin{center}
\resizebox{0.8\linewidth}{!}{
\begin{tabular}{lcccccc}
\toprule
 & 
\textbf{NRT} & \textbf{Att2Seq} & \textbf{CER} & \textbf{PETER} & \textbf{PEPLER} & \textbf{XRec} \\
\midrule
\multicolumn{7}{c}{\textbf{Toys}} \\
\midrule
  \text{StEnt-P} & \worst{0.0466\sd{0.1706}} & \best{0.0916\sd{0.1542}} & 0.0805\sd{0.1663} & \secondbest{0.0831\sd{0.1715}} & 0.0729\sd{0.1655} & \secondworst{0.0538\sd{0.1178}} \\
  \text{StEnt-R} & \worst{0.0127\sd{0.0567}} & \best{0.0532\sd{0.1131}} & 0.0522\sd{0.1180} & \secondbest{0.0530\sd{0.1181}} & 0.0441\sd{0.1104} & \secondworst{0.0435\sd{0.1033}} \\
  \text{StEnt-F1} & \worst{0.0061\sd{0.0328}} & \best{0.0410\sd{0.0907}} & 0.0349\sd{0.0880} & \secondbest{0.0360\sd{0.0902}} & 0.0299\sd{0.0831} & \secondworst{0.0259\sd{0.0674}} \\
\cmidrule(lr){2-7}
  \text{StCoh-P} & \secondworst{-0.0261\sd{0.2527}} & 0.0048\sd{0.2388} & \secondbest{0.0160\sd{0.2343}} & \best{0.0204\sd{0.2381}} & 0.0098\sd{0.2334} & \worst{-0.0803\sd{0.2594}} \\
  \text{StCoh-R} & \worst{-0.1639\sd{0.2025}} & -0.0662\sd{0.2217} & \secondbest{-0.0590\sd{0.2077}} & -0.0649\sd{0.2125} & \secondworst{-0.0895\sd{0.2131}} & \best{-0.0588\sd{0.2014}} \\
\midrule
\multicolumn{7}{c}{\textbf{Clothes}} \\
\midrule
  \text{StEnt-P} & \secondbest{0.2217\sd{0.2074}} & \secondworst{0.1777\sd{0.2106}} & 0.2110\sd{0.2329} & 0.2202\sd{0.2444} & \best{0.2422\sd{0.2254}} & \worst{0.1407\sd{0.1679}} \\
  \text{StEnt-R} & \secondbest{0.1284\sd{0.1803}} & 0.1141\sd{0.1690} & \secondworst{0.1102\sd{0.1694}} & \worst{0.1099\sd{0.1682}} & \best{0.1402\sd{0.1892}} & 0.1160\sd{0.1686} \\
  \text{StEnt-F1} & \secondbest{0.1259\sd{0.1630}} & \secondworst{0.1016\sd{0.1507}} & 0.1077\sd{0.1563} & 0.1088\sd{0.1590} & \best{0.1381\sd{0.1708}} & \worst{0.0895\sd{0.1280}} \\
\cmidrule(lr){2-7}
  \text{StCoh-P} & 0.1222\sd{0.3323} & \secondworst{0.0931\sd{0.2965}} & 0.1188\sd{0.3254} & \secondbest{0.1341\sd{0.3319}} & \best{0.1539\sd{0.3222}} & \worst{0.0250\sd{0.2749}} \\
  \text{StCoh-R} & \secondbest{0.0372\sd{0.2375}} & -0.0108\sd{0.2761} & \secondworst{-0.0112\sd{0.2433}} & \worst{-0.0156\sd{0.2448}} & \best{0.0390\sd{0.2522}} & 0.0169\sd{0.2640} \\
\midrule
\multicolumn{7}{c}{\textbf{Beauty}} \\
\midrule
  \text{StEnt-P} & \secondworst{0.1367\sd{0.3265}} & 0.1555\sd{0.2022} & 0.1980\sd{0.2693} & \best{0.2076\sd{0.2751}} & \secondbest{0.2001\sd{0.2447}} & \worst{0.1162\sd{0.1641}} \\
  \text{StEnt-R} & \worst{0.0218\sd{0.0713}} & \best{0.0915\sd{0.1404}} & \secondworst{0.0718\sd{0.1308}} & 0.0755\sd{0.1314} & \secondbest{0.0764\sd{0.1330}} & 0.0755\sd{0.1282} \\
  \text{StEnt-F1} & \worst{0.0323\sd{0.1039}} & \best{0.0813\sd{0.1297}} & 0.0753\sd{0.1351} & 0.0806\sd{0.1406} & \secondbest{0.0810\sd{0.1345}} & \secondworst{0.0597\sd{0.1004}} \\
\cmidrule(lr){2-7}
  \text{StCoh-P} & 0.0963\sd{0.3575} & \secondworst{0.0823\sd{0.2744}} & 0.1326\sd{0.3352} & \best{0.1474\sd{0.3401}} & \secondbest{0.1373\sd{0.3146}} & \worst{-0.0140\sd{0.2975}} \\
  \text{StCoh-R} & \worst{-0.2454\sd{0.1970}} & \secondbest{-0.0298\sd{0.2420}} & \secondworst{-0.0746\sd{0.2295}} & -0.0722\sd{0.2297} & -0.0570\sd{0.2227} & \best{-0.0286\sd{0.2322}} \\
\midrule
\multicolumn{7}{c}{\textbf{Sports}} \\
\midrule
  \text{StEnt-P} & 0.1588\sd{0.3428} & \secondworst{0.1521\sd{0.1912}} & \secondbest{0.2063\sd{0.2794}} & \best{0.2117\sd{0.2868}} & 0.1574\sd{0.2211} & \worst{0.0973\sd{0.1679}} \\
  \text{StEnt-R} & \worst{0.0215\sd{0.0660}} & \best{0.0763\sd{0.1303}} & 0.0706\sd{0.1300} & 0.0704\sd{0.1305} & \secondbest{0.0744\sd{0.1336}} & \secondworst{0.0488\sd{0.1048}} \\
  \text{StEnt-F1} & \worst{0.0326\sd{0.0991}} & 0.0675\sd{0.1161} & \best{0.0709\sd{0.1349}} & \secondbest{0.0706\sd{0.1353}} & 0.0643\sd{0.1204} & \secondworst{0.0386\sd{0.0874}} \\
\cmidrule(lr){2-7}
  \text{StCoh-P} & 0.0794\sd{0.3867} & \secondworst{0.0760\sd{0.2571}} & \secondbest{0.1541\sd{0.3338}} & \best{0.1590\sd{0.3403}} & 0.0989\sd{0.2761} & \worst{-0.0673\sd{0.3415}} \\
  \text{StCoh-R} & \worst{-0.3174\sd{0.1908}} & \best{-0.0256\sd{0.2180}} & -0.0783\sd{0.2248} & -0.0814\sd{0.2268} & \secondbest{-0.0366\sd{0.2100}} & \secondworst{-0.1245\sd{0.2909}} \\
\midrule
\multicolumn{7}{c}{\textbf{Cellphones}} \\
\midrule
  \text{StEnt-P} & \best{0.2480\sd{0.3170}} & \secondworst{0.1096\sd{0.1697}} & 0.1878\sd{0.2654} & 0.1617\sd{0.2582} & \secondbest{0.2238\sd{0.2545}} & \worst{0.1036\sd{0.1664}} \\
  \text{StEnt-R} & \worst{0.0367\sd{0.1009}} & \secondbest{0.0629\sd{0.1235}} & 0.0577\sd{0.1220} & \secondworst{0.0509\sd{0.1135}} & \best{0.0661\sd{0.1326}} & 0.0514\sd{0.1128} \\
  \text{StEnt-F1} & \worst{0.0399\sd{0.1061}} & 0.0463\sd{0.0977} & \secondbest{0.0580\sd{0.1234}} & 0.0494\sd{0.1143} & \best{0.0703\sd{0.1328}} & \secondworst{0.0402\sd{0.0906}} \\
\cmidrule(lr){2-7}
  \text{StCoh-P} & \best{0.1570\sd{0.4285}} & \secondworst{0.0157\sd{0.2578}} & 0.1078\sd{0.3437} & 0.0782\sd{0.3346} & \secondbest{0.1365\sd{0.3595}} & \worst{-0.0641\sd{0.3295}} \\
  \text{StCoh-R} & \worst{-0.1805\sd{0.2257}} & \best{-0.0575\sd{0.2377}} & -0.1101\sd{0.2336} & \secondworst{-0.1259\sd{0.2222}} & -0.0692\sd{0.2234} & \secondbest{-0.0656\sd{0.2322}} \\
\bottomrule
\end{tabular}
}
\label{tab:NLI-Statement-level-results}
\end{center}
\vspace{-6mm}
\end{table*}
\subsection{NLI-based Statement Evaluation (StEnt/StCoh)}
Table~\ref{tab:NLI-Statement-level-results} presents results for our NLI-based metrics, which compute entailment and contradiction scores over statement pairs. Overall scores are low, corroborating LLM-based findings on limited factual consistency while inducing different system rankings.
\\

\noindent
\textbf{Entailment.}
Precision scores (\texttt{StEnt-P}) range from $4.66\%$ (NRT on \textit{Toys}) to $24.80\%$ (NRT on \textit{Cellphones}), showing substantial cross-dataset variability. Precision can be inflated for models generating few statements when those statements are common in the data.
Recall scores (\texttt{StEnt-R}) support this interpretation: despite higher precision, NRT achieves only $3.67\%$ recall on \textit{Cellphones}. Recalls remain generally low, with a maximum of $14.02\%$ for PEPLER on \textit{Clothes}.
%
\\

\noindent
\textbf{Coherence.}
The \texttt{StCoh-*} metrics assess coherence via entailment-contradiction differences. Negative values indicate more contradiction than entailment. Even recent models like XRec yield negative precision and recall on several datasets, with only a few models (e.g., PEPLER on \textit{Clothes}) achieving positive but modest scores. These results highlight that beyond low factual consistency, models can produce statements that directly contradict the reference.
\\

\noindent
\textbf{LLM-based vs NLI-based Results.}
Overall, we observe a disparity between our LLM-based and NLI-based metrics, attributable to their differing granularities. 
NLI metrics compare statement pairs, while LLM-based metrics compare each statement against the full explanation. 
Although NLI models enable efficient pairwise comparison at scale, prohibitively expensive for LLMs, the LLM-based approach better preserves the complete reference context when scoring individual statements. 
We also observe that the results of LLM-based metrics are more consistent and stable than those of NLI-based metrics.
Nevertheless, both metric families converge on the same conclusion: models exhibit poor factual consistency.

\section{Discussion}
\label{sec:discussion}
Our experimental results reveal several critical insights about the state of factual consistency in text-based explainable recommendation systems. 

\noindent
\textbf{The Factuality Gap.}
Our experiments uncover a dramatic disconnect between surface-level text quality and factual accuracy. Models achieve impressively high scores on standard similarity metrics suggesting near-human quality text generation. However, when evaluated through our statement-level factual consistency metrics, these same models exhibit strikingly poor performance.
It suggests that models have learned to generate fluent, contextually appropriate text that \textit{appears} explanatory but frequently fails to ground its claims in verifiable evidence.

\noindent
\textbf{Precision versus Recall Trade-offs.}
A consistent pattern across our results is that models struggle with both precision and recall, though often in different ways. The low precision scores indicate frequent hallucination of explanatory content not supported by the evidence. The even lower recall scores reveal that models fail to recover most of the ground-truth explanatory passages, leaving critical aspects of the user's preferences unaddressed.



\noindent
\textbf{Limitations.}
Our work has several limitations that suggest directions for future research:
\textit{(1) LLM-Based Extraction.} Our statement extraction pipeline relies on LLMs, which may introduce errors or biases. While we employ carefully designed prompts and  regular manual qualitative verifications, the extraction process is not perfect yet. Future work could explore more robust extraction methods, potentially incorporating human verification for critical applications.
\textit{(2) Granularity of Ground-truth.} Our rule-based aggregation of statements into explanations preserves all extracted content but may not reflect the natural structure or emphasis users would prefer. Alternative approaches, perhaps learning to select and organize statements based on user preferences or contextual relevance, 
could yield more realistic ground-truth.

\section{Conclusion}
This paper presents a comprehensive investigation into the factual consistency of text-based explainable recommendation systems, revealing a critical gap between surface-level text quality and factual accuracy. Through the introduction of a statement-level evaluation framework, augmented benchmark datasets, and novel factuality metrics, we have demonstrated that current state-of-the-art models, despite achieving impressive fluency scores, frequently hallucinate explanatory content and fail to ground their outputs in verifiable evidence.
This disconnect underscores a fundamental limitation in current evaluation practices, which prioritize semantic similarity over factual grounding. 
Our findings suggest that achieving factual consistency in explainable recommendation will require fundamental innovations in model architectures, training objectives, and evaluation. 

\bibliographystyle{splncs04}
\bibliography{output}

\end{document}